\documentclass[aps,pra,prl,twocolumn,superscriptaddress,showpacs]{revtex4}
\usepackage{graphicx}
\usepackage{amsfonts,amsmath,amssymb}
\usepackage{dcolumn}
\usepackage{bm}

\begin{document}
\def\bbox#1{\hbox{\boldmath${#1}$}}
\def\blambda{{\hbox{\boldmath $\lambda$}}}
\def\eeta{{\hbox{\boldmath $\eta$}}}
\def\bxi{{\hbox{\boldmath $\xi$}}}

\title{
Quantum Treatment of Continuum Electrons in the Fields of
Moving Charges}
\author{Teck-Ghee~Lee} \email[Corresponding author email:
]{leetg@ornl.gov}

\affiliation{Physics Division, Oak Ridge National Laboratory, Oak
Ridge, TN 37831}

\affiliation{Department of Physics and Astronomy, University of
Kentucky, Lexington, KY 40506}

\author{S. Yu. Ovchinnikov}

\affiliation{Physics Division, Oak Ridge National Laboratory, Oak
Ridge, TN 37831}

\affiliation{Department of Physics and Astronomy, University of
Tennessee, Knoxville, TN 37496}

\author{J. Sternberg}

\affiliation{Department of Physics and Astronomy, University of Tennessee, Knoxville, TN 37496}

\author{V. Chupryna}

\affiliation{Department of Physics and Astronomy, University of Tennessee, Knoxville, TN 37496}

\author{D. R. Schultz}

\affiliation{Physics Division, Oak Ridge National Laboratory, Oak
Ridge, TN 37831}

\author{J. H. Macek}

\affiliation{Physics Division, Oak Ridge National Laboratory, Oak
Ridge, TN 37831}

\affiliation{Department of Physics and Astronomy, University of
Tennessee, Knoxville, TN 37496}

\date{\today}

\begin{abstract}
An {\it ab initio}, three-dimensional quantum mechanical
calculation has been performed for the time-evolution of continuum
electrons in the fields of moving charges. Here the essential singularity
associated with the diverging phase factor in the continuum wave function
is identified and removed analytically. As a result, the continuum components of
the regularized wave function are slowly varying with time. Therefore, one can propagate continuum electrons to asymptotically
large times and obtain numerically stable, well-converged ejected electron
momentum spectra with very low numerical noise. As a consequence, our
approach resolves outstanding controversies concerning structures
in electron momentum distributions.  The main conclusions are general and are illustrated
here for ionization of atomic hydrogen by proton impact. Our results show that in order to obtain correct long-time free-particle propagation, the essential singularity identified here should be removed from the continuum components of solutions to the time-dependent
Schr\"{o}dinger equation.

\end{abstract}


\maketitle

Tracing the evolution of continuum electrons in the presence of
time-dependent external fields as they move from microscopic to
macroscopic distances has been of great fundamental interest since
the beginning of quantum mechanics \cite{Sakurai}. Observations
are made at distances where they can be detected by devices such
as photon or particle detectors which also have macroscopic
dimensions.  Substantial progress has been made in the
measurements of continuum electrons owing to the development of
the cold target recoil ion momentum spectroscopy technique
\cite{Dorner, Moshammer,Schulz, Ullrich}, for example. We describe
a corresponding theoretical advance in this Letter. This advance
is made possible by combining two lines of research, namely, the
lattice time-dependent Schr\"{o}dinger equation (LTDSE) method of
Schultz and coworkers \cite{Wells, Schultz1, Schultz2} and the
removal of essential singularities of continuum wave packets
described in this Letter but implicit in the dynamical Sturmian
theory of Ovchinnikov and coworkers \cite{Macek}.

Although standard numerical techniques successfully apply for
time-dependent Schr\"{o}dinger equation (TDSE) of bound states,
they meet with many difficulties for the continuum components due
to highly oscillatory phase factors $\exp[ir^2/2t]$ as is evident
from the free-particle propagator (see Eq.(2.5.16) in
Ref.\cite{Sakurai}). Because these phase factors have essential
singularities for large distances, both fundamental and numerical
difficulties arise. Even with today's most powerful computers,
propagation to large times without accurate treatment of the
essential singularity is not possible. Here we present a technique
to remove the singularity and apply it to a prototypical system
treated in an explicitly time-dependent approach. To the best of
our knowledge, no standard numerical methods exist that can
rigorously compute the time-evolution of quantum systems from
atomic to macroscopic scales.

With removal of the essential singularity, accurate numerical
solution of the TDSE to asymptotically large times becomes
possible. This is important for physical processes where electron
motion at large times, in contrast to just near the parent nuclei,
plays a decisive role.  For example, in ion-atom collisions,
classical trajectory Monte Carlo calculations~\cite{Carlos} show
that cusps in the ejected electron spectra do not emerge until the
target and projectile species are separated by more than 5,000
a.u., many times larger than the initial atomic dimensions.
Moreover,  cusps are known on fundamental grounds to occur and are
prominent features of all ionization processes where nuclear
charges are not completely screened in the final state. For
ionization by strong laser fields, rescattering of electrons from
atomic or molecular species is important for multiple ionization
and high harmonic generation~\cite{Yakovlev}. Such rescattering
occurs due to oscillations of the laser field so that the electron
periodically turns around at large distances and recollides with
the core of the species from which it originated.

Thus, the key innovation illustrated in the present work is of
far-reaching importance because the motion of continuum electrons
in external fields is central to many processes which have only
been treated classically until now. Specifically, we show how to
treat these processes on an {\it ab initio}, completely quantum
mechanical basis. To begin with, the electronic wave function is
written in the form (atomic units are used throughout the paper
unless otherwise indicated)
\begin{equation}
\label{eq1} \Psi({\bbox r}, t) = \chi({\bbox q},
\theta)\Phi({\bbox q},\theta),
\end{equation}
where ${\bbox q}$=$({\bbox r} v \sin\theta)/\Omega$ and
$t$=$-(\Omega \cot\theta)/v^2$, $\theta$ varies from $0$ to $\pi$,
and $v$ and $\Omega$ are arbitrary parameters in general but later
$v$ will represent a physical velocity. The factor
\begin{equation}
\label{eq2} \chi({\bbox q}, \theta) = \left(
\frac{i\sin\theta}{\Omega}\right)^{3/2} \exp[-i (\Omega q^2
\cot\theta)/2]
\end{equation}
is highly oscillatory and has an essential
singularity at $1/q$=0 ({\it i.e.}~$q \rightarrow \infty$).

The {\em regularized} wave function $\Phi(\bbox q,\theta)$
however, has slow-varying continuum components and satisfies a
regularized time-dependent Schr\"{o}dinger equation (RTDSE)
\begin{equation}
\left[ { - \frac{1}{2}\nabla^2 _{\bbox q} + \frac{1}{2}\Omega
^2q^2 + V({\bbox q},\theta )} \right]\Phi ({\bbox q},\theta )
 = i\Omega\frac{\partial \Phi ({\bbox q},\theta )}{\partial \theta },
 \label{eq3}
\end{equation}
where $V({\bbox q},\theta)$ is an external, possibly
time-dependent, potential. Time-dependent scale-transformations
have been used to solve a model problem exactly~\cite{Solovev} and
a one dimension problem numerically~\cite{Sidky1}, but the method
introduced here enables efficient and full numerical calculations.

In the present application where the bare projectile of charge
$Z_P$ collides with a one-electron target of nuclear charge $Z_T$,
the interaction is given by
\begin{eqnarray}
V({\bbox q},\theta) = \frac{\Omega}{v\sin\theta}\left(-{Z_P\over
|{\bbox q} + {\bbox Q(\theta)/2}|}
 - {Z_T\over |{\bbox q}-{\bbox Q}(\theta)/2|} \right)
\end{eqnarray}
where  ${\bbox Q(\theta)}$ can represent any trajectory.  Here we
use a straight-line trajectory to describe the nuclear motion,
{\it i.e.} ${\bbox Q(\theta)} = \cos\theta\hat{{\bbox v}}+
vb\sin\theta \hat{{\bbox b}}/ \Omega$, where $v$ is now the
relative velocity and $b$ is the impact parameter.

The initial regularized wave function is given by
\begin{eqnarray}
\Phi_0({\bbox q},\theta_0)=\frac{\Psi_0(v^{-1}{\bbox
q}~\Omega\csc\theta_0,-v^{-2}\Omega\cot\theta_0)}{\chi({\bbox q},
\theta_0)},
\end{eqnarray}
where $\Psi_0({\bbox r},t_0)$ is the full wave function at the
initial time $t_0=-(\Omega\cot\theta_0)/v^2$.  At asymptotically
large times as $\theta \rightarrow \pi$, the continuum components
of  $\Phi(\bbox q,\theta)$ are independent of $\Omega$ and
directly give the ejected electron momentum
distributions~\cite{Macek}
\begin{equation}
A({\bbox k})=\lim_{\theta \rightarrow \pi} \Phi ({\bbox
q},\theta), \label{eq6}
\end{equation}
where ${\bf q}\rightarrow{\bbox k}/v$ in the limit and {\bbox k}
is the electron wave vector. Note that the present formulation
does not require projection onto dynamic two-center continuum
states at large times since the bound states shrink to vanishingly
small regions in ${\bbox q}$-space. Eq.~(\ref{eq3}) can be then
solved using a number of different numerical methods.

We compute the ejected electron spectra for proton impact
ionization of hydrogen atoms in the impact energy range of 5$-$25
keV where there are long standing controversies regarding the
momentum distribution of ejected
electrons~\cite{Ullrich,Sidky2,Schultz1}. Owing to its numerical
accuracy and efficiency we adopt here the LTDSE technique
\cite{Wells, Schultz2, Schultz1}. As the two moving charges become
separated by a very large distance, the electronic probability
density in the continuum becomes infinitesimally small, spread
over an enormous spatial volume. As a result, previous
calculations have always terminated at relatively small distances
on the order of 30 to 100 a.u. \cite{Sidky2,Schultz2,Schultz1}
before the continuum electron density becomes so small that it
poses intractable numerical difficulties.

Because of the present singularity-free, scaled representation,
the RTDSE does not suffer from this impediment. Furthermore, due
to the steep walls of the harmonic oscillator potential in
Eq.~(3), reflections at the lattice boundaries are inherently
insignificant. Finally, removing the essential singularity also
allows us to integrate along a path in the complex $\theta-$plane
where the interaction potential never vanishes for real {\bbox q},
allowing us to avoid conventional, {\it ad hoc} softcore
modifications of the potentials near the Coulomb singularities.

\begin{widetext}

\begin{figure} [ht]
\begin{center}
\includegraphics[angle=0,scale=0.65]{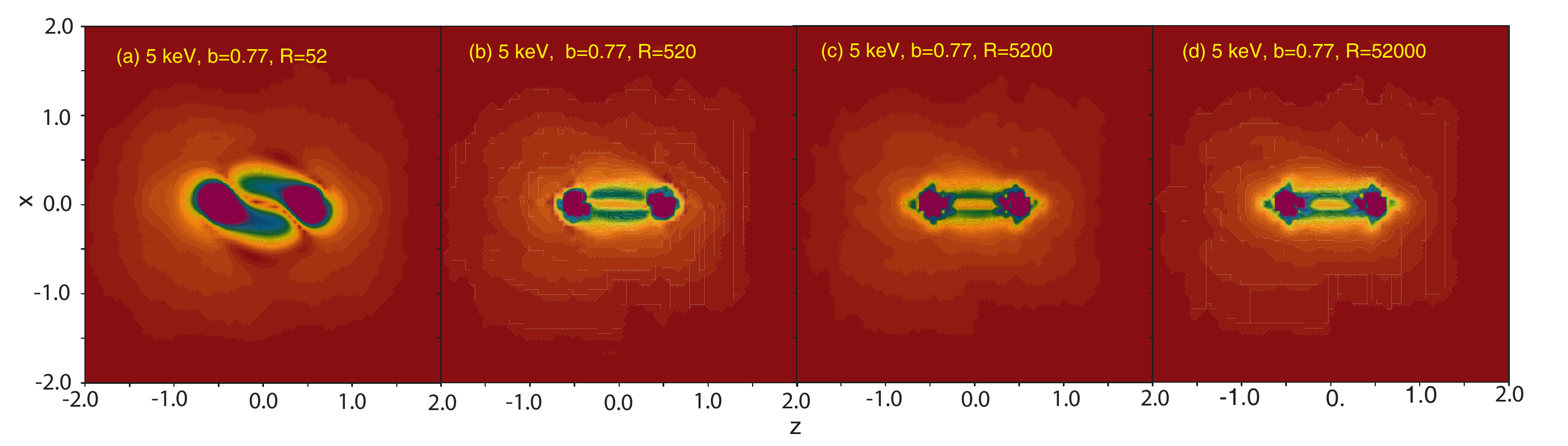}
\caption{The time-evolution of the spectrum of ejected electrons at a collision
energy of 5 keV and an impact parameter of $b$=0.77 a.u. for proton
impact ionization of hydrogen. The time integration is
terminated at 4 internuclear separations: (a) $R$=52 a.u.,
(b) $R$=520 a.u., (c) $R$=5,200 a.u. and (d) $R$=52,000 a.u. The electronic amplitude (Eq.~(\ref{eq6})) has values of $\sim$0.02 near the edges of the distribution and $\sim$0.64 a.u. at $q_x$=0 and $q_z$=$\pm 0.5$.  Note that the electron wave vector $\bm{k} $ is measured in units of the relative ion velocity $v$. \label{fig1}}
\end{center}
\end{figure}

\end{widetext}

Regarding specific parameters of the present calculations, we
choose $\Omega\approx 1$ and the scaled box size $|q_i| \leq 7, $
$i=x, y, z$ with 256$^3$ points in order to minimize the level of
numerical noise in the resulting electron momentum distributions.
This is judged by performing a test calculation with projectile
charge set to zero. Using the quantity $1-
|\langle\Phi(\theta)|\Phi_0(\theta)\rangle|^2$ as measure of the
noise, we find it to be less than $3 \times 10^{-7}$. This error
is small enough so that all of the momentum distribution within a
box $|k/v| < 2.5$ ({\it e.g.}, see Fig.\ref{fig1}) is numerically
significant.  Such small and controllable errors set a standard
for accuracy of full 3-dimensional numerical solutions of the TDSE
and allow us to see features of electron distributions that are
beyond the previous state-of-the-art.

We begin by noting that in this energy range considered here,
ionization occurs via transfer of energy from nuclear to
electronic motion in regions where there are periodic, unstable
classical trajectories. There are two types of these trajectories,
namely, the superpromotion (S) trajectories along the internuclear
axis ({\it e.g.}, as in the Fermi Shuttle mechanism
\cite{Sulik}), and the top-of-barrier (T) trajectories where the
electron moves perpendicular to the axis joining the nuclei near
the region where the net force on the electron vanishes. The
S-type motion is effective as the nuclei approach while the T-type
is effective as they recede. Both types of trajectories are
related to quantal transitions between H$_2^{+}-$like adiabatic
molecular energy curves~\cite{Pieksma}.

Figure \ref{fig1} shows 2-dimensional slices of the ejected electron momentum (or wave
vector $\bm{k}$) distributions taken with the $q_z$-axis parallel to
the projectile velocity $\bm{v}$, the $q_x$-axis parallel to the
impact parameter $\bm{b}$ and $y$ equal to zero, computed for collision
energy 5 keV with $b$=0.77 a.u., and with the time integration
terminated at four internuclear separations: $R$ = 52, 520,
5,200 and 52,000 a.u. One can see that the distribution changes
significantly between 52 and 5,200 a.u., predicted in a special limiting case \cite{Serge2}, with a small, but noticeable, change
between 5,200 and 52,000 a.u.

\begin{figure}
\includegraphics[scale=0.5]{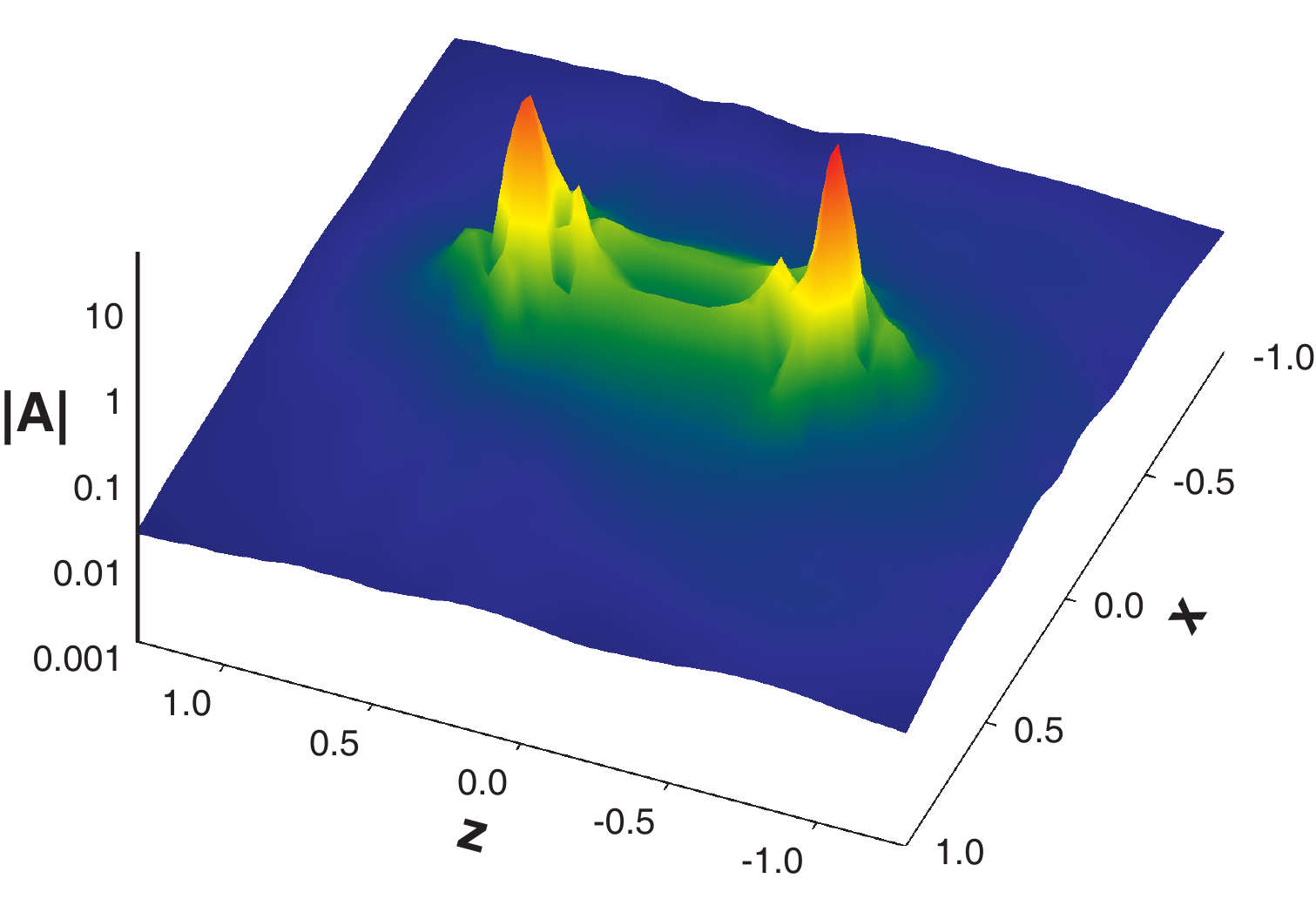}
\caption{Plot of the ionization amplitude $|A|$ of Eq.~(\ref{eq6})  showing target and  projectile cusps with the bound states removed.
\label{fig2}}
\end{figure}

\begin{figure}
\includegraphics[scale=0.5]{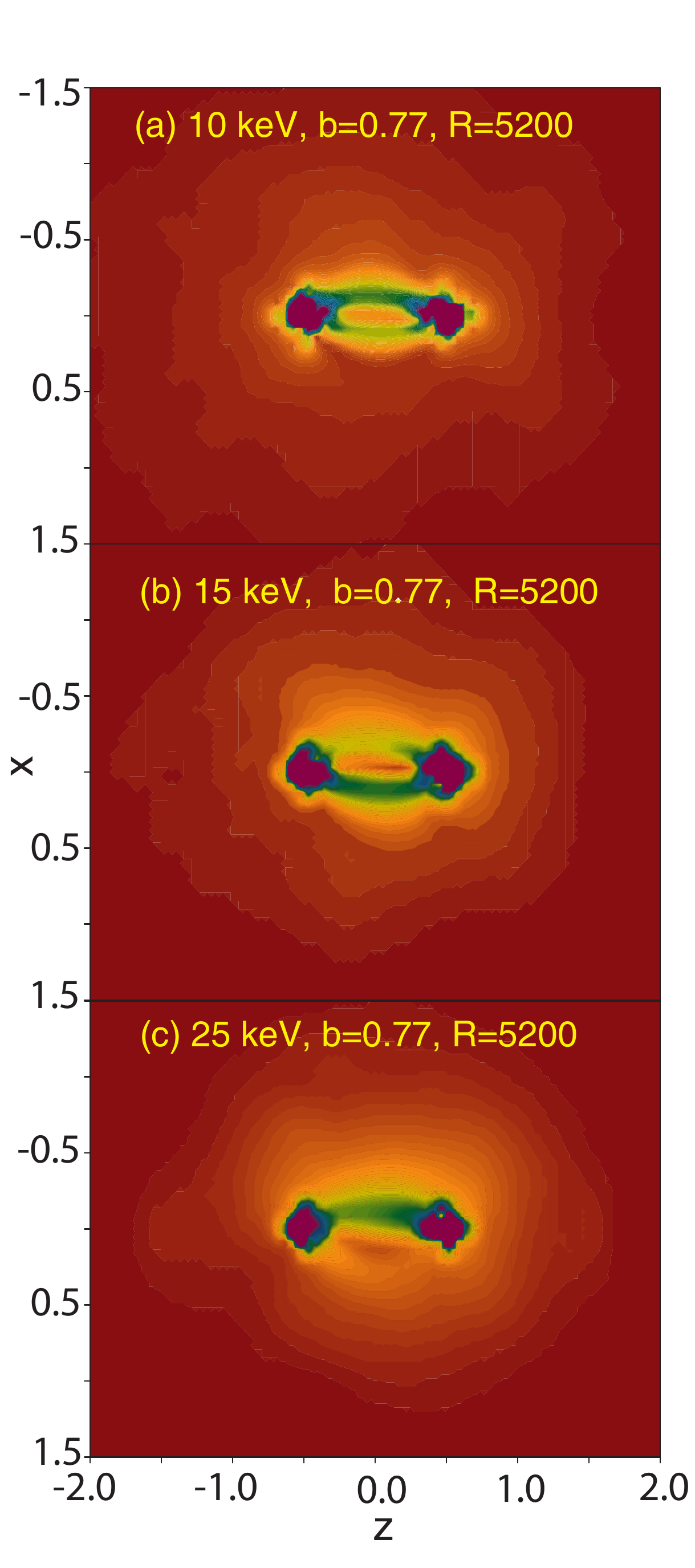}
\caption{Same as in Fig.~\ref{fig1} but for higher collision energies with $R$=5,200 a.u .
The amplitude ranges from $\sim$0.04 near the edges of the distributions to $\sim$1.36 a.u. at $q_x$=0 and $q_z$=$\pm 0.5$.
} \label{fig3}
\end{figure}

At all post-collisional internuclear distances there are fast electrons, {\it i.e.}~electrons with velocities greater than the projectile velocity $v$,
distributed approximately isotropically about the center of mass.
These electrons originate from two  S-type transitions
near $R$=0.4 a.u. and $R$=1 a.u. The electrons are distributed in angle
according to the angular part of the adiabatic H$_2^{+} $ wave
function at complex values of the internuclear distance $R$~\cite{Serge2}.
The relevant angular parts have not been reported in the literature, but
they are assumed to be given in terms of spherical harmonics with
relatively low values of angular momentum $\ell$. This agrees qualitatively with
the {\it ab initio} distributions reported here.

The `elliptical' ring of electrons between the target and
projectile is another remarkable feature which becomes narrower as
$R$ increases, and can be interpreted in terms of T-type
transitions, {\it i.e.} electrons are ejected by a process that
involves a united-atom rotational coupling transition followed by
T-promotion to  the continuum. It is known that united-atom
rotational coupling produces a state of $\pi$ symmetry with a node
along the internuclear axis. The shape of this feature at $R$=52
a.u. and its narrowing for increasing $R$ support this
interpretation.

Electron cusps at $\bm{k}$=$\pm
\hat{\bm{v}}/2$ are also seen in Fig.~\ref{fig1}, although one
must remember that the bound state electrons are not separated
from the cusps in this figure.  It is necessary to extrapolate
the distribution to small $\bm{k}$ when $R$ is asymptotically
large in order to identify these features correctly. Figure~\ref{fig2}
shows the ionization amplitude $|A|$ at $R$=52,000 a.u.
The electrons in the cusps are a small fraction of
the total that are ejected, however, even this small
component can be identified with the theoretical approach that we
report here for the first time with such a time-dependent grid approach.

The identification of the these features is supported by the
distributions shown in Fig.~\ref{fig3}  for increasing projectile velocity.
Note that the superpromotion features shrink in extent but do not
change shape.  This is in accord with predictions of  an exponential $\exp[-0.4(k/v)^2 v]$
distribution~\cite{Macek}.  Alternatively, the T-electron
distribution retains the same spatial extent but oscillates as identified in Ref. \cite{Macek}.
In contrast with that earlier approximate theory, it can be seen
that some of the electrons actually move from the center to the target and projectile nuclei.
Note that all of the features identified here are superimposed coherently so
that interference effects may emerge.  Indeed, interference structure near the
cusps in Fig.~\ref{fig2} is apparent.   A more detailed analysis of the time-dependent
wave functions shows that the S-type electrons interfere with the T-type electrons that
have moved from the midpoint of the internuclear axis.  Such interference has
not been noted previously since a coherent treatment of S- and T-type
processes has been beyond the reach of theory.

The most important conclusion from the present calculations is
that, even for the many-facited electron distributions seen in
Figs.~\ref{fig1}-\ref{fig3}, in experiment, and in earlier
calculations \cite{Macek}, the overall phase factor of the
electronic wave function in continuum is given by Eq.~(\ref{eq2}).
In $\{\bm{r}, t\}$-space this factor for $R>>1$ is just
$\exp[ir^2/2t]$ as derived from the free electron propagator. With
this phase factor removed, a regularized Schr\"{o}dinger equation
is obtained that is solved efficiently using the LTDSE method.
This has allowed us to develop a general theoretical approach that
allows one to solve the time-dependent Schr\"{o}dinger equation
from microscopic to macroscopic distances accurately. The method
is well-suited for treating continuum electrons in time-dependent
fields. Computations of ejected electron distributions produced by
proton impact on atomic hydrogen have been used here to illustrate
the method. The results show measurable features not amenable to
previous theoretical approaches. The features are in qualitative
accord with electron distributions measured for proton impact on
helium \cite{Dorner}.

This research is sponsored by the Office of Basic Energy Sciences,
U.S. Department of Energy, under Contract No. DE-AC05- 96OR22464
through a grant to Oak Ridge National Laboratory, which is managed
by UT-Battelle, LLC under Contract No. DE-AC05-00OR22725. JHM, SYO
and JS, acknowledge support by DOE grant DE-FG02-02ER15283 to the
University of Tennessee. TGL acknowledges support by NASA grant
NNG05GD81G to the University of Kentucky.

\end{document}